\documentclass[pre,aps,amsfonts,amssymb,amsmath,showpacs]{revtex4}

\usepackage{ifpdf}
\ifpdf
\usepackage{subfigure}
\usepackage[pdftex]{graphicx}
\else
\usepackage{graphicx}
\fi

\usepackage{setspace}
\usepackage{psfrag}
\usepackage{epstopdf}
\usepackage{graphicx}
\DeclareGraphicsExtensions{.eps}

\usepackage{amsmath,graphicx,epsfig,bm,color}
\usepackage{amssymb,amsfonts}
\usepackage{rotating}
\usepackage{graphics}
\usepackage{setspace}

\definecolor{red}{rgb}{1,0,0}
\definecolor{orange}{rgb}{1,0.5,0}
\definecolor{blue}{rgb}{0,0,1}
\definecolor{green}{rgb}{0,1,0}

\begin{document}
\title{\bf Self-similar formation of an inverse cascade in vibrating elastic plates.}
\author{Gustavo D\"uring${}^{1}$, Christophe Josserand${}^{2}$ and Sergio Rica${}^{3}$}
\affiliation {
${}^1$Facultad de F\'isica, Pontificia Universidad Cat\'olica de Chile,
Casilla 306, Santiago, Chile\\
${}^2$Sorbonne Universit\'es, CNRS \& UPMC Univ Paris 06, UMR 7190, Institut d'Alembert, F-75005, Paris, France\\
${}^3$Facultad de Ingenier\'ia y Ciencias, Universidad Adolfo Ib\'a\~{n}ez, Avda. Diagonal las Torres 2640, Pe\~{n}alol\'en, Santiago, Chile\\
}
\date{\today}

\begin{abstract}
The dynamics of random weakly nonlinear waves is studied in the framework of vibrating thin elastic plates. Although it has been previously predicted that no stationary inverse cascade of constant wave action flux could exist in the framework of wave turbulence for elastic plates, we present substantial evidence of the existence of a time dependent inverse cascade, opening up the possibility of self organization for a larger class of systems. This inverse cascade transports the spectral density of the amplitude of the waves from short up to large scales, increasing the distribution of long waves despite the short wave fluctuations. This dynamics appears to be self-similar and possesses a power law behavior in the short wavelength limit which is significantly different from the exponent obtained via a Kolmogorov dimensional analysis argument. Finally,  we show explicitly a tendency to build a long wave coherent structure in finite time. 
\end{abstract}

\maketitle

\section{ Introduction} Oscillating random waves are present in a myriad of situations in nature, displaying a large variety of scales and exhibiting 
turbulent-like behavior, the so-called wave turbulence~\cite{ZakharovBook,NazarenkoBook,NewellRumpf}. Of particular interest are the oscillations over the surface of the sea, Rossby waves in atmospherical science, nonlinear optics, the plasma oscillations and the vibration of elastic bodies such as piano strings, timbals, or more complex singing 
bowls, bells or gongs. Because of the intrinsic nonlinearity of the basic underlying physics of these problems, and because of the randomness of the phase of the 
oscillations, only a statistical description seems reasonable. The weak turbulence theory provides such a statistical description for the asymptotic  long time behavior of  the spectral wave amplitude, in the case where nonlinearities are small. In particular, it describes the energy transfer among the different modes in agreement with the conservation of the total energy of the waves. 
More precisely, this wave turbulence theory provides kinetic equations for the long-time evolution of the spectral amplitude for dispersive wave systems \cite{ZakharovBook,NazarenkoBook,NewellRumpf}.  In the present context of small nonlinearities, we will use interchangeably wave turbulence theory and weak turbulence theory and refer to it as WTT. Remarkably, such kinetic equations exhibit stationary solutions corresponding to equipartition or constant flux cascades of the energy, namely the Rayleigh-Jeans solution and the Kolmogorov-Zakharov (KZ) spectrum respectively. The search of the KZ spectra has motivated exhaustive studies these last fifty years~\cite{NewellRumpf}, regaining recent interest, because of the parallel development of new theoretical and experimental findings. Among them, we mention the cases of surface capillary waves \cite{falcon,falconduring}, surface gravity waves \cite{Zak2004,lukashuk} and elastic waves of thin plates \cite{during,arezki,mordant08} for instance. 
While such dynamics corresponds usually to a direct cascade of energy towards the small scales, the formation of large scale structure can
sometimes be observed.
This is the case in particular for Bose-Eintein condensation~\cite{lacaze,lowtemp,natalia,becnls} or in two dimensional hydrodynamic turbulence \cite{kraichnan}, 
where the self organization process is a consequence of an inverse cascade. This inverse cascade transfers some quantity (particles, enstrophy, wave action) from the small scales toward the large scales leading to the formation of coherent structures. The formation of an inverse cascade in different systems has always been related to the existence of a conserved quantity at least at the weakly non linear level. 
For instance, when using the Gross-Pitaevskii or nonlinear Schr\"odinger equation to model the Bose-Einstein condensates, WTT predicts the existence of an inverse cascade of mass (a conserved quantity). Similarly, in the case of surface gravity waves, an inverse cascade  of the conserved wave action is deduced and numerically observed \cite{shrira,korotkevich}. Finally, we want to emphasize that besides the description of the stationary solutions of the dynamics, the kinetic equations that are deduced by the WTT give a very good framework to investigate non stationary situations involved in wave
systems such as transitory or decaying regimes that often lead to self-similar dynamics~\cite{falko91,CNP03,Luc}.
The goal of this paper is to show, using the elastic vibrating plate, that the formation of an inverse cascade does not generally require a conserved quantity, opening up the possibility of self organization for a larger class of systems. In particular, nothing prevents the existence of a time dependent inverse cascade that would 
transfer wave action from short scale to large scale.

The paper is organized as follows: Section \ref{sec:wtt}, introduces the dynamical version of the F\"oppl--von K\'arm\'an equations which provide the basic nonlinear equations for vibrating elastic plates, containing both bending and stretching. Then we summarize the main findings of the WTT of a vibrating plate, in particular the concept of the Kolmogorov-Zakharov spectra. Section \ref{sec:inverse} presents the numerical simulations of a vibrating plate which is forced only at very short wavelength, displaying a striking inverse cascade of wave action. Section \ref{sec:blowup} analyzes the numerical evidence of a self similar evolution which suggests a blows-up in finite time. Finally, we conclude with an overall discussion of the problem. 

\section{ Wave turbulence theory of vibrating elastic plates} 
\label{sec:wtt}

\subsection{The F\"oppl--von K\'arm\'an equations for elastic plates}
Vibrating elastic plates offer, perhaps, the most suitable weakly non linear wave system. 
It is studied within the framework of the dynamical version of the F\"oppl--von K\'arm\'an equations
\cite{landau} which model the dynamics of the  out of plane displacements of the plate. We shall use the same notations as in Ref. \cite{during},  but we write them in dimensionless units. We choose $ l={h}/{\sqrt{3(1- \sigma^2)}}$ as  the unit of length, and $l \sqrt{\frac{\rho}{E}}$ as the unit of time. Here $h$ is the thickness of the elastic sheet, the material has a mass density $\rho$, a Young modulus $E$ and its Poisson ratio is $\sigma$. 
In these units the equations read:
\begin{eqnarray} 
\frac{\partial^2 \zeta}{\partial t^2} &=& -\frac{1}{4}\Delta^2\zeta +
\{\zeta,\chi\}  ;
\label{foppl0}\\
\Delta^2\chi &=&- \frac{1}{2}\{\zeta,\zeta\}.
\label{foppl1}
\end{eqnarray}
The out of plane displacement of the plate in physical units is thus $l \zeta(x,y,t)$, and the Airy stress function is $E l^2 \chi(x,y,t)$. Equation (\ref{foppl1}) for  the Airy stress function
$\chi(x,y,t)$ may be seen as the compatibility equation for the in--plane stress tensor which follows the dynamics at the lowest order \footnote{In the derivation we have omitted the inertia of the in-plane modes of oscillations, or in other words we assume that the in-plane displacements are negligible and  the static equilibrium holds, so that, as said, equation (\ref{foppl1}) describes the dynamics.}.
The characteristic size of the plate is $L$, thus the dynamics of a free plate is governed by a single dimensionless parameter: $ \Lambda =\frac{ L}{l}=\sqrt{3(1- \sigma^2)} \frac{L}{h}$, which is typically of the order of $10^3$ up to $10^4$. $\Delta=\partial_{xx}+\partial_{yy}$ is the usual Laplacian and the bracket 
$\{\cdot,\cdot\}$ is defined by $\{f,g\}\equiv f_{xx}g_{yy}+f_{yy}g_{xx}-2f_{xy}g_{xy},$  which is an exact divergence, so that equation (\ref{foppl0}) preserves the momentum of the center of mass, hence, $\partial_{tt} \int \zeta(x,y,t) \, dx\,dy=0$. Moreover, the total energy:
\begin{equation}
E = \int \left(\frac{1}{2}(\partial_t\zeta)^2+ \frac{1}{8} (\Delta \zeta )^2  -  \frac{1}{2}(\Delta\chi)^2 - \frac{1}{2} \chi \{\zeta,\zeta\} \right)\,dx\,dy\, ,
\label{totalenergy}
\end{equation} is also conserved by the dynamics (\ref{foppl0},\ref{foppl1}).
Finally, small plane waves perturbations ($\zeta \sim e^{i( {\bm k}\cdot {\bm x} -\omega_k t)}$ with ${\bm x} =(x,y)$) of a plane plate are dispersive 
with the usual ballistic behavior of bending waves, that is $\omega_{ k} =  \frac{1}{2} |\bm k|^2$~\cite{landau}.

\subsection{Wave turbulence equations for the spectral densities}

As already discussed in Ref. \cite{during}, equations (\ref{foppl0},\ref{foppl1}) exhibit a Hamiltonian structure which is easy revealed in Fourier space, defined by $\zeta_{\bm k}(t)= \frac{1}{2\pi} \int \zeta({\bm x},t) e^{i  {\bm k}\cdot {\bm x} } d^2{\bm x}$, with $\zeta_{\bm k}= \zeta_{-{\bm k}}^*$. The Hamiltonian structure allows one to performs a canonical transformation
\begin{equation}  \zeta_{\bm k} =  \frac{1}{\sqrt{2\omega_k}} ( A_{\bm k}+ A^*_{-{\bm k}} )\label{Canonical}
\end{equation}
  which lead to a diagonalized form of the  wave equation:   \begin{equation}\frac{d A_{\bm k}}{d t}+i\omega_{ k} A_{\bm k}= i N_3( A_{\bm k}),  \label{EqnForAs}
  \end{equation}where $N_3(\cdot)$ abbreviates  the cubic nonlinear interaction term given explicitly in Ref. \cite{during}.

The WTT describes the long time statistical behavior of  weakly non linear random waves. The analysis is based on an infinite hierarchy of integro-differential equations for the cumulants of the canonical variables which maybe deduced directly from (\ref{EqnForAs}). In the weak wave amplitude limit,
a multiscale asymptotic expansion of these hierarchy of equations provides a rational scheme for solving every cumulants~\cite{ZakharovBook,NazarenkoBook,NewellRumpf}. As a result, the second order cumulant
\begin{equation}
\left< A_{ \bm k_1} A^{*}_{{\bm k}_2} \right> = n_{ {\bm k}_1} \delta^{(2)}(k_1+k_2),\label{spectrum}
\end{equation}
is shown to control the long-time dynamics of the wave system, where $n_{\bm k}$ is called the spectrum of the wave. 
Other second order cumulants vanishes in the weak  amplitude (long time) limit  \cite{ZakharovBook,NazarenkoBook,NewellRumpf}, in particular
$$\left< A_{ \bm k_1} A_{{\bm k}_2} \right>\to 0 \quad \& \quad \left< A^*_{ \bm k_1} A^*_{{\bm k}_2} \right>\to 0. $$ 

The asymptotic perturbation scheme of this theory provides at first order a nonlinear frequency shift to the linear waves, leading to an effective oscillation frequency $\omega_k^{eff}=\omega_k + \omega_k^{(1)} +\dots$. This correction due to weak nonlinear effects is a function of the mean spectral density $n_{\bm  k}(t)$ (\ref{spectrum}) and it reads (in the dimensionless units) \cite{during2}~:
  \begin{equation}\omega^{(1)}_k=    \frac{\pi}{ 2 }    \left[ \int_0^k \frac{q^2}{k^2} n_q \, q d{q}  +  \int_k^\infty \frac{k^2}{q^2} n_q \, q d{q} \right] .
  \label{OmegaRenorm}
  \end{equation}
Notice that this frequency correction was also obtained by considering a limited number of nonlinear interactions~\cite{japs2}.
In addition, in the WWT this frequency shift (\ref{OmegaRenorm}) is useful to quantify the nonlinear effects, and by consequence the validity of the WTT. Indeed, the 
ratio $\omega^{(1)}_k/\omega_k $ indicates the relative importance of the nonlinear term with respect to the linear behavior. The uniform validity of the WTT 
requires that this ratio should satisfy $|\omega^{(1)}_k/\omega_k |\ll1$ for all the wave numbers. If this number is of the order of unity, wave turbulence is no longer 
valid, at least for the concerned scales.

At the next order, WTT  provides a kinetic equation that governs the mean spectral density evolution $n_{\bm  k}(t)$, which reads \cite{during}:
\begin{eqnarray}
\frac{d}{dt}n_{k}&=& {\mathcal C}[n_k] = 12\pi \int d{\bm{k}}_{1}d{\bm{k}}_{2}d{\bm{k}}_{3}\, \left|{ J_{-{\bm k}, {\bm k}_1;{\bm k}_2,{\bm k}_3}  }\right|^2   \nonumber\\ & \times&  \sum_{s_1s_2s_3}    n_{{k}_1}n_{{k}_2}n_{{k}_3}n_{{k}} 
\left(\frac{1}{n_{{k}}}-\frac{s_1}{n_{{k}_1}}-\frac{s_2}{n_{{k}_2}}-\frac{s_3}{n_{{k}_3}}\right)
\delta^{(2)} ({\bm k}-{\bm k}_1-{\bm k}_2-{\bm k}_3) 
\delta( \omega_{{k}}-s_1\omega_{{k}_1}-s_2\omega_{{k}_2}-s_3\omega_{{k}_3}).
\label{KineticEqn}
\end{eqnarray} 

The  coefficient $ J_{-{\bm k}, {\bm k}_1;{\bm k}_2,{\bm k}_3}  $,  in Eq. (\ref{KineticEqn}), comes from the fourth order nonlinearities in the total energy (\ref{totalenergy}), and they are given explicitly in Ref. \cite{during}. The details of this scattering function $J$ is not needed here and we omit to write it for the sake of simplicity, but, for the purpose of this work, we only have to notice that it has 
zero degree of homogeneity in $k$, that is   
$$J_{-\lambda{\bm k} ,\lambda{\bm k}_1;\lambda{\bm k}_2,\lambda{\bm k}_3} = J_{-{\bm k}, {\bm k}_1;{\bm k}_2,{\bm k}_3}. $$
Finally, we mention that this function $J_{-{\bm k}, {\bm k}_1,{\bm k}_2,{\bm k}_3}$ vanishes as $\bm k\rightarrow 0$, so that the spectrum, does 
not vary at $k=0$, in agreement with the original plate equations (\ref{foppl0}) and (\ref{foppl1}). If $n_{k=0}=0$ at $t=0$, then the spectrum vanishes 
at $k=0$ for all time. 

The spectral dynamic described by the kinetic equation (\ref{KineticEqn})  corresponds to four-waves resonances  which enforce the energy and momentum 
conservations in each interaction. However, in contrast with the case of diluted gases where each collision preserves the number of particles, the total number of 
waves involved in the interaction is not formally conserved for vibrating plates.  Moreover, it is interesting to notice that while the total energy (\ref{totalenergy}) is 
conserved by the original F\"oppl--von K\'arm\'an equations (\ref{foppl0},\ref{foppl1}), the kinetic equation preserves only the quadratic part of the total energy, 
namely, $E_2 = \int \left(\frac{1}{2}(\partial_t\zeta)^2+ \frac{1}{8} (\Delta \zeta )^2 \right)\,dx\,dy$. 

The sum in (\ref{KineticEqn}) rules for $s_i = \pm 1$, that is, the kinetic equation (\ref{KineticEqn}) contains $8$ terms. 
Among them, the one corresponding to all $s_i$ equal is not resonant and it thus vanishes. Three other terms (all identical by symmetries) correspond to 
interactions of two waves coming-in and two waves going-out, so that the total number of waves is preserved by this interaction. We will refer to these terms further 
on as the $2\leftrightarrow 2$ resonant case. Finally, there are four other interaction terms corresponding to  one(three) wave(s) coming-in and  three(one) waves going-out, 
which are referred as the $3\leftrightarrow 1$ resonant case. In this later case, the total number of waves is not preserved formally by the four waves 
interaction. We denote these two different interaction terms by $ {\mathcal C}_{22} $ and $ {\mathcal C}_{13}$, respectively, so that $ {\mathcal C}[n_k]=  {\mathcal C}_{22}[n_k]+  {\mathcal C}_{13}[n_k]$.
This non wave action conservation represents a major difference with most of the known four-wave interaction systems such as surface  gravity 
waves~\cite{zakgrav66}, or nonlinear optics \cite{Dyachenko-92}. To our knowledge the only known physical systems that exhibit these two kind of 
interactions ($2\leftrightarrow 2$ and $3\leftrightarrow 1$) are the symmetric capillary waves at the interface between two fluids \cite{falconduring} and  the elastic 
plates \cite{during}.

In conclusion, although the kinetic energy $ { \mathcal E} = \int { \omega_k }  n_{\bm  k}(t) d^2k,$ is preserved by the dynamics (\ref{KineticEqn}), the wave action, ${\mathcal N}= \int  n_{\bm  k}(t) d^2k,$ is not.

\subsection{Kolmogorov-Zakharov spectra}

Although the wave action is not preserved by the dynamics, local conservation equations can be deduced from the kinetic equation. Indeed,
the change in time of the energy spectral density $E(k)= 2\pi k \omega_k n_k $, can be written, after (\ref{KineticEqn}), as
\begin{equation}
\frac{d }{dt} E(k) = -\frac{d }{dk}  P(k) \quad {\rm where}\quad P(k) = 2 \pi \int_k^\infty \omega_q {\mathcal C}[n_q]\, q\, dq
\label{fluxEdef}
\end{equation}
is the energy flux, which depends, in principle, explicitly on the wavenumber $k$ and $t$. Similarly, the wave action flux $Q(k)$ maybe defined via the wave action spectral density: $N(k)= 2\pi k n_k $, through: 

\begin{equation}
\frac{d }{dt} N(k)= \frac{d  }{dk} Q(k) \quad {\rm with}\quad Q(k) = 2 \pi \int_0^k {\mathcal C}[n_q]\, q\, dq.
\label{fluxNdef}
\end{equation}
As the energy flux, the wave action flux may depend on the wave number and time. Notice that the special writing of equations (\ref{fluxNdef}) may induce the wrong impression that $\int N(k) dk$ is conserved by the dynamics, but this is not so, because $Q(0) \neq Q(\infty)$. Wave turbulence theory predicts a class of exact power-law solutions of the kinetic equation (\ref{KineticEqn}), found by Zakharov \cite{ZakharovBook}, which keep the fluxes constant. 
More precisely a (direct) energy cascade is found for which  the energy flux $P$ is constant. Similarly, if the wave action is conserved by the dynamics, then, an (inverse) wave action cascade corresponding to a constant 
wave action flux $Q$ can be exhibited. These solutions are named the Kolmogorov-Zakharov (KZ) spectra, because Zakharov's findings are in close relation with the Kolmogorov scaling arguments used in fluid turbulence.

Taking an arbitrary power law solution, $n_k = A\, k^{-2 x}$, introducing this into the collisional operator ${\mathcal C}[n_k]$ one readily gets:
$${\mathcal C}[A\, k^{-2 x}] = A^3 I(x) k^{2-6x}.$$
Here $I(x)$ is a pure function, which depends only on the exponent $x$, and whose expression has been explicitly  written in Ref. \cite{during}.
Following (\ref{fluxEdef}) and (\ref{fluxNdef}), the fluxes are then given by:
\begin{equation} P = A^3 \frac{\pi I(x)}{6(x-1)} k^{6(1-x)} \quad {\rm and}\quad Q  = A^3  \frac{\pi I(x)}{(2-3 x)} k^{4 - 6 x}.
\label{Fluxes}
\end{equation}
Constant energy or wave action fluxes are obtained if the exponents take the values $x=1$ or $x=2/3$ respectively. Such arguments guarantee only that
the scaling of the solution is consistent with the collisional operator. However, since the denominator vanish for those exponents,  one needs in addition that the collisional operator vanishes in order to obtain stationary
solutions of the kinetic equation.
This condition is not satisfied for elastic plates, where the inverse cascade of wave action, $x=2/3$ is not a root of $I(x)$. Only the terms corresponding the  $2\leftrightarrow 2$ resonances vanish for
$x=2/3$, while the terms due to $3\leftrightarrow 1$ resonances do not. Then, the  flux $Q$ formally diverges and the KZ solution is not valid.

On the other hand,  $I(x)$ has a double root for $x=1$ due to a special degeneracy ($I(x)$ vanishes quadratically near $x=1$, $I(x)\sim (1-x)^2$). It indicates that both the Rayleigh-Jeans and the KZ solutions exist for $x=1$. In practice, the resulting flux is zero, thus a  logarithmic  correction should be included on the final spectrum \cite{during}.
Therefore, the above considerations imply that only one cascade is guaranteed, namely a direct cascade of energy toward the small scales, which remarkably is not a simple power law. This KZ spectrum, predicted by~\cite{during}, reads (the numerical pre-factor is discussed in Ref. \cite{during2}):

\begin{equation}
n^{direct}_{ k} \sim  P^{1/3 }  \frac{\ln^{1/3}(k_*/k)}{k^2},
\label{KZ}
\end{equation} 
where $P$ is the energy flux, and $k_*$ a cutoff scale. These solutions have been observed in numerical simulations performed with an ad-hoc dissipation concentrated at small scales only~\cite{during,japs,mordant13}.
However, experimental observations~\cite{arezki,mordant08,mordant09,mordant11} present a slightly different behavior for the direct energy cascade, which is understood as follows. In Ref. ~\cite{EPL}, it is shown that the dominant low frequency dissipation rate of the damping suppresses  the existence of a window of transparency in the wave number range probed by the experiment. It turns out, that the stationary spectrum for a vibrating plate comes from the balance between the kinetic collision integral, the forcing and  damping, displaying  no simple KZ scaling spectrum, and  exhibiting a strong dependence on the  damping mechanisms \cite{EPL}.

We conclude this section with the following remark. If one neglects the $3\leftrightarrow 1$ resonances in the kinetic equation, that is if one imposes $\mathcal{C}_{13}[n_k]\approx 0$, then an inverse cascade of wave action with constant flux can exist in addition to the direct cascade of energy. This inverse cascade of wave action (from small scales to large 
scales) reads:  
\begin{equation}
n^{inverse}_{ k}\sim   Q^{1/3 }  \frac{1}{k^{4/3}},
\label{KZinverse}
\end{equation}
where $Q<\infty$ is identified as the (constant) wave action flux.
This inverse cascade transfers wave action between modes, a self-organization process which may lead to the formation of coherent structures and eventually 
to the breakdown of the WTT~\cite{Dyachenko-92,becnls,becnls2}. 
 
The goal of this paper is therefore to investigate the plate dynamics by forcing the vibrations at small scales only in order to observe the genuine transfer of wave 
action towards large scales despite the presence of the $3\leftrightarrow 1$ interactions. 

\section{Manifestation of an Inverse Cascade} \label{sec:inverse}

We solve numerically the coupled set of dynamical equations (\ref{foppl0}, \ref{foppl1}) using a pseudo-spectral method which takes advantage of the linear wave dynamics in Fourier space. 
Formally, the equations (\ref{foppl0},\ref{foppl1}) read, in Fourier space:
\begin{equation}\ddot \zeta_{\bm k} = - \omega_k^2 \zeta_{\bm k}  + NL_{\bm k} - D_{\bm k} \dot\zeta_{\bm k} + I_{\bm k},
\label{PlateEqnFourier}
\end{equation}
where $NL_k$ stands for the Fourier transform of the nonlinear term including equation (\ref{foppl1}), $D_k$  represents a linear damping and $I_k$ the forcing in 
spectral space. Finally, the temporal integration is performed in the Fourier space using a second order Adams-Bashford scheme. Standard dealiaising technique 
for cubic nonlinearity has been tested in previous works~\cite{during} with no qualitative changes in the results so that no dealiaising is used in the present 
simulations.
 \begin{figure}[h!]
\begin{center}
\centerline{a) \includegraphics[width=5cm]{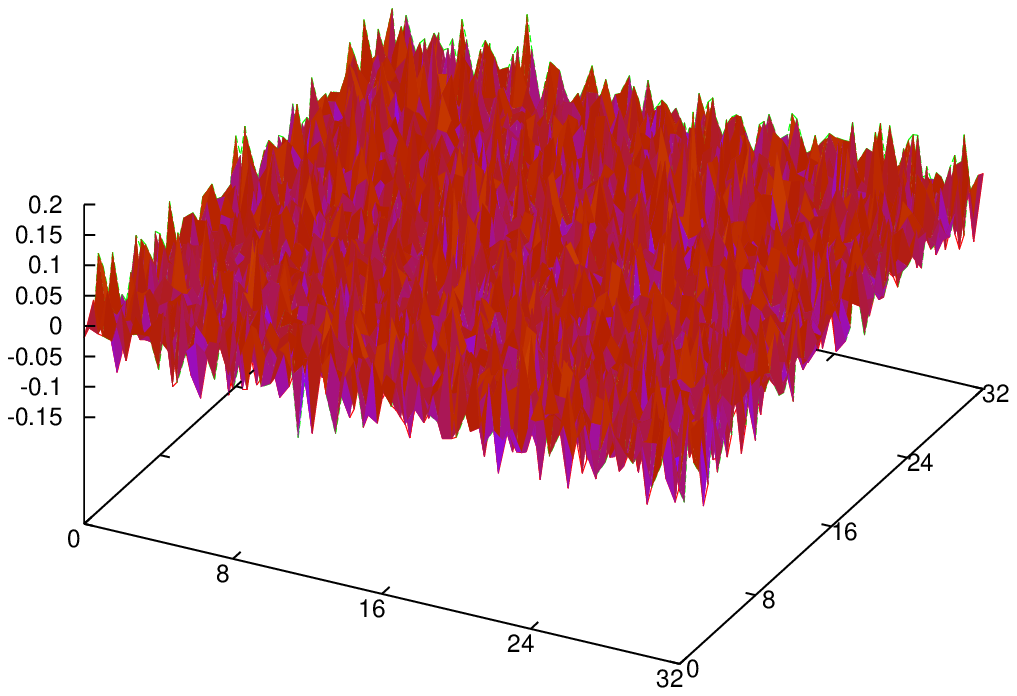} \quad b)  \includegraphics[width=5cm]{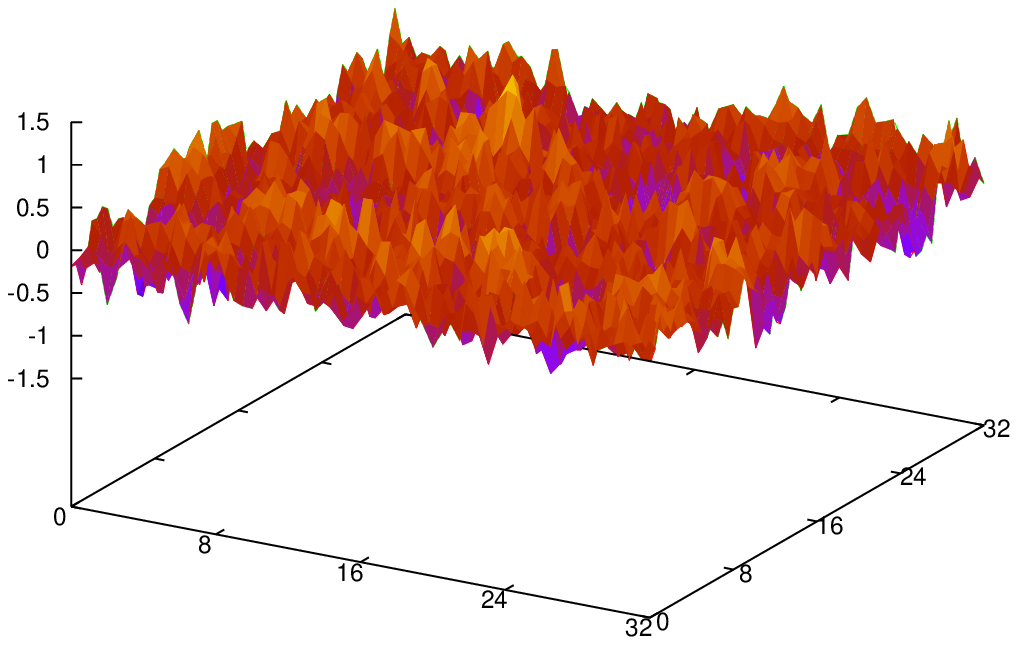} \quad c)  \includegraphics[width=5cm]{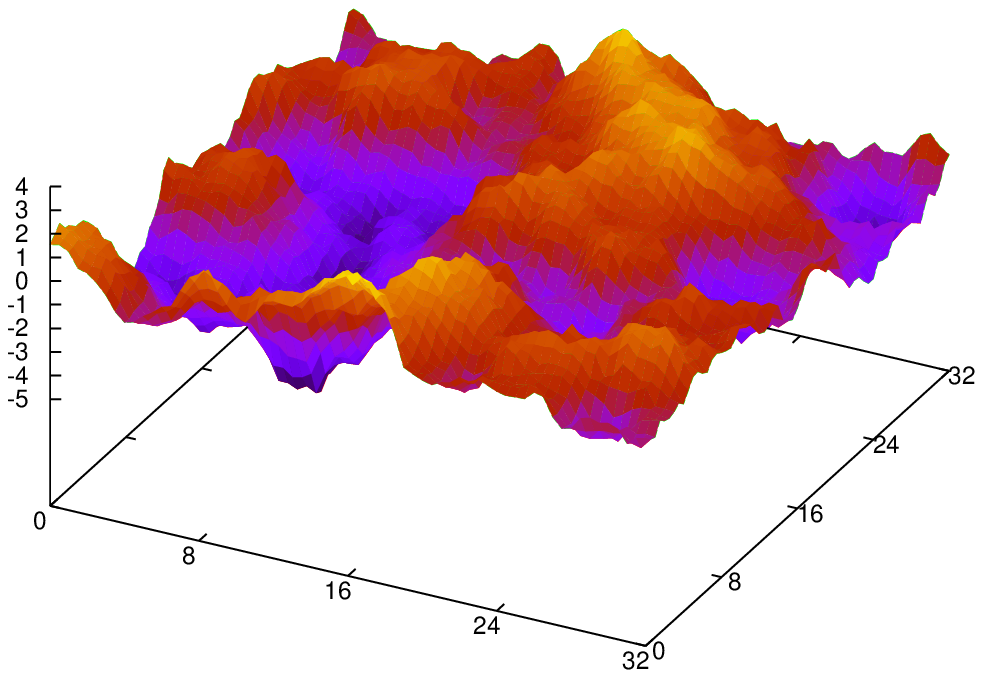}  }
\centerline{d) \includegraphics[width=5cm]{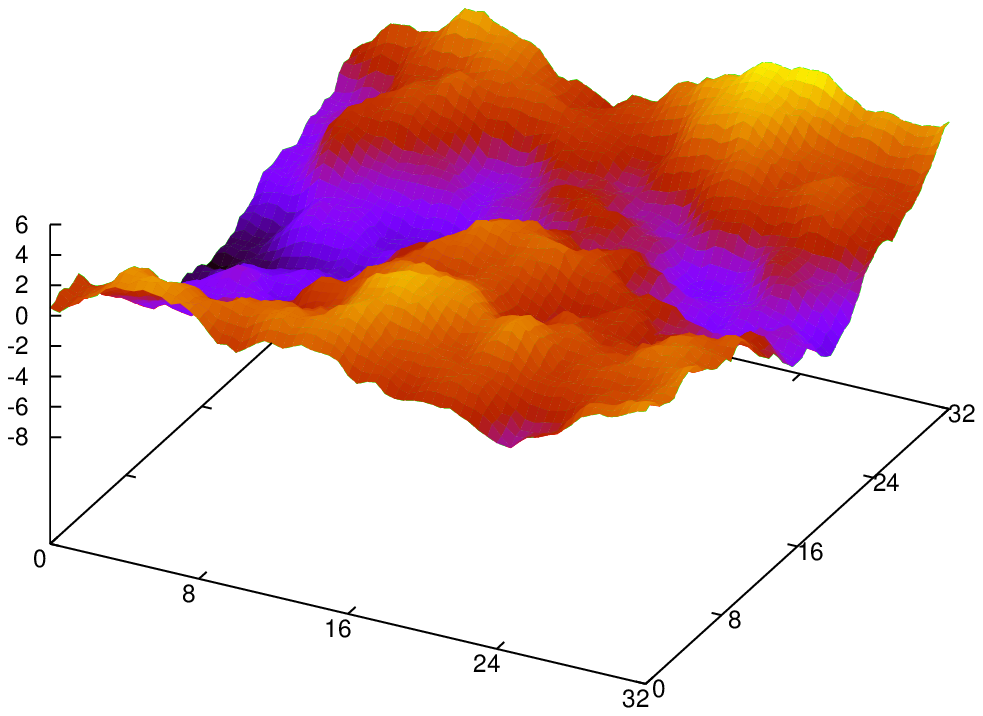} \quad e)  \includegraphics[width=5cm]{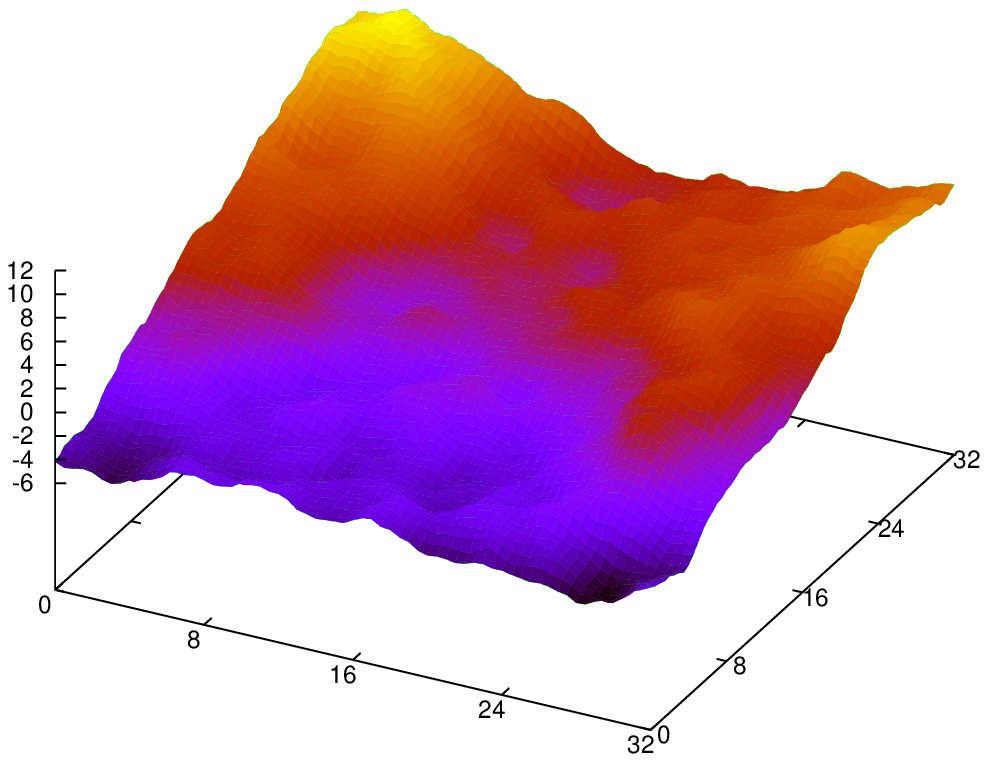} \quad f)  \includegraphics[width=5cm]{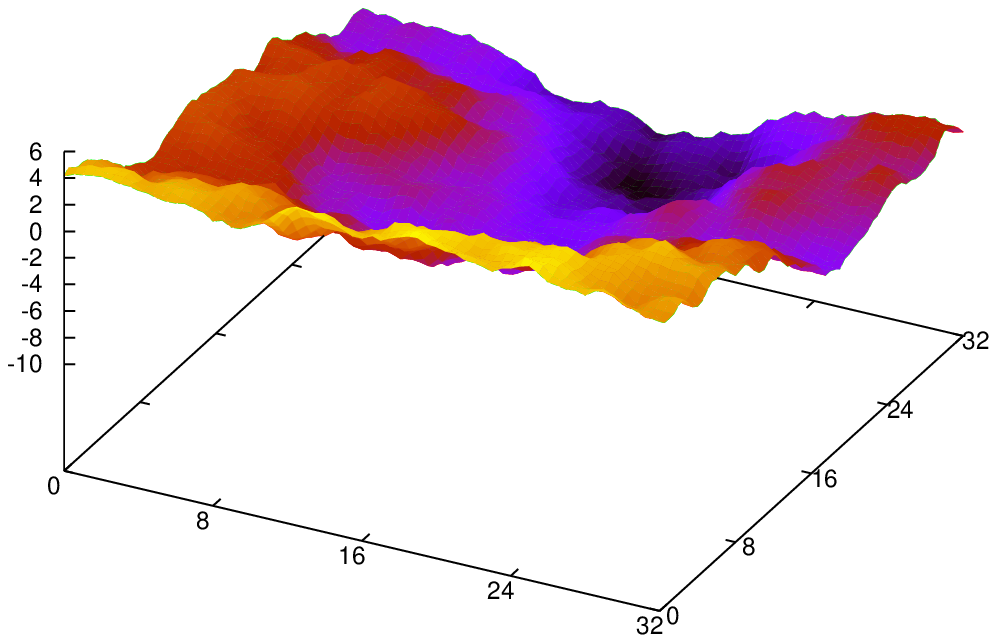}  }
\caption{Snapshots the out of plate displacement $\zeta(x,y,t)$ obtained by the numerical simulation of the F\"oppl--von K\'arm\'an equations (\ref{foppl0},\ref{foppl1}) at six different times : a) $t=22\times 10^3$; b) $t=50\times 10^3$; c) $t=62\times 10^3$; d) $t=82\times 10^3$; e) $t=162\times 10^3$; and f) $t=202\times 10^3$. The injection is made at small scales, with $k_i=4.5$ and $\delta_i=0.5$. Numerical dissipation acts at smaller scale, starting at the end of the injection range ($k_d=5$). The amplitude of the injection is $A_i=0.0001$. The system size is $1024^2$ units with $2048^2$  modes and $k_c = 2 \pi$. Note that  the vertical scales differ from one figure to the other, growing from a) to f). Despite this change of scale, the amplitude of the small spatial scales can always be observed. Notice the formation of a large scale structure as time increases.}
\label{snapshots}
\end{center}
\end{figure}

In the present work we use periodic boundary conditions, which are the natural framework to investigate the features of the wave turbulence, the number of modes 
ranging from $512^2$ up to  $2048^2$, with a mesh size $dx=1/2$, leading to the spectral ultraviolet cut-off $k_c = \pi/dx= 2 \pi$. For numerical stability, the time
step used for the simulations is $dt=0.02$ unit time.
To observe the dynamics towards large scale, we force and dissipate the system at small scales only. The dissipation  is given by  $D_{\bm k}=-\eta (k^2-k_d^2) H(k-k_d)$ where  $\eta$ is the amplitude of the damping, $H(\cdot)$ is the Heaviside function and $2\pi/k_d$ the characteristic scale below which the only dissipation acts. The forcing will be non zero only in the finite range $\left[k_i-\delta_i,k_i+\delta_i \right]$ where
 $$I_{\bm k}=A_i
    \frac{(k^2-(k_i-\delta_i)^2)((k_i+\delta_i)^2-k^2)}{k_i^4}e^{ i \theta_{\bm k}(t)}. $$  Here $k_i$ is  the characteristic scale of the forcing, $\delta_i$ and $A_i$ its the width and amplitude respectively. The angular variable $\theta_k(t)$ is a random phase taken in the interval $[0,2\pi]$,  which also changes randomly in time. Notice that this process injects both energy and wave action around $k_i$ since one cannot separate them formally. 
To illustrate this inverse transfer mechanism, we take $k_d=k_i+\delta_i$ so that the inertial range for an energy cascade vanishes and the low frequency (large scale) inverse transparency window is the largest possible available. Finally, we have checked numerically that the results do not depend on the details 
of the dissipation at small scales.

Although it is required to dissipate the energy at small scales (large $k$) to
reach numerically a stationary state, we have realized distinct numerical simulations with a sink and without it located near $k=0$, and we conclude that it is not 
required to absorb nor dissipate the energy (nor the wave action) at the large scale (small $k$) to reach numerically a stationary state in the time scale of the 
simulations.  This can be explained firstly by the fact that energy is eventually dissipated at small scales leading to the general balance between the injected and
the dissipated energy. For the wave action, since this quantity is not conserved by the dynamics, everything works as if the wave action is formally absorbed by a sink at
$k=0$ (which is a neutral mode) so that there is no need to add such an absorption term near $k=0$ in the dynamics.

Fig. \ref{snapshots} shows the snapshots of the plate deformation at six distinct times of the evolution. Notice the apparent formation of a coherent structure 
which at the end oscillates at the largest possible mode. This coherent structure appears as a consequence of the long time evolution which is 
mostly characterized by the largest modes of oscillation of a plate 
plus small fluctuations. 
\vskip 1 cm
\begin{figure}[h]
\begin{center}
\centerline{a)\, \includegraphics[width=8cm]{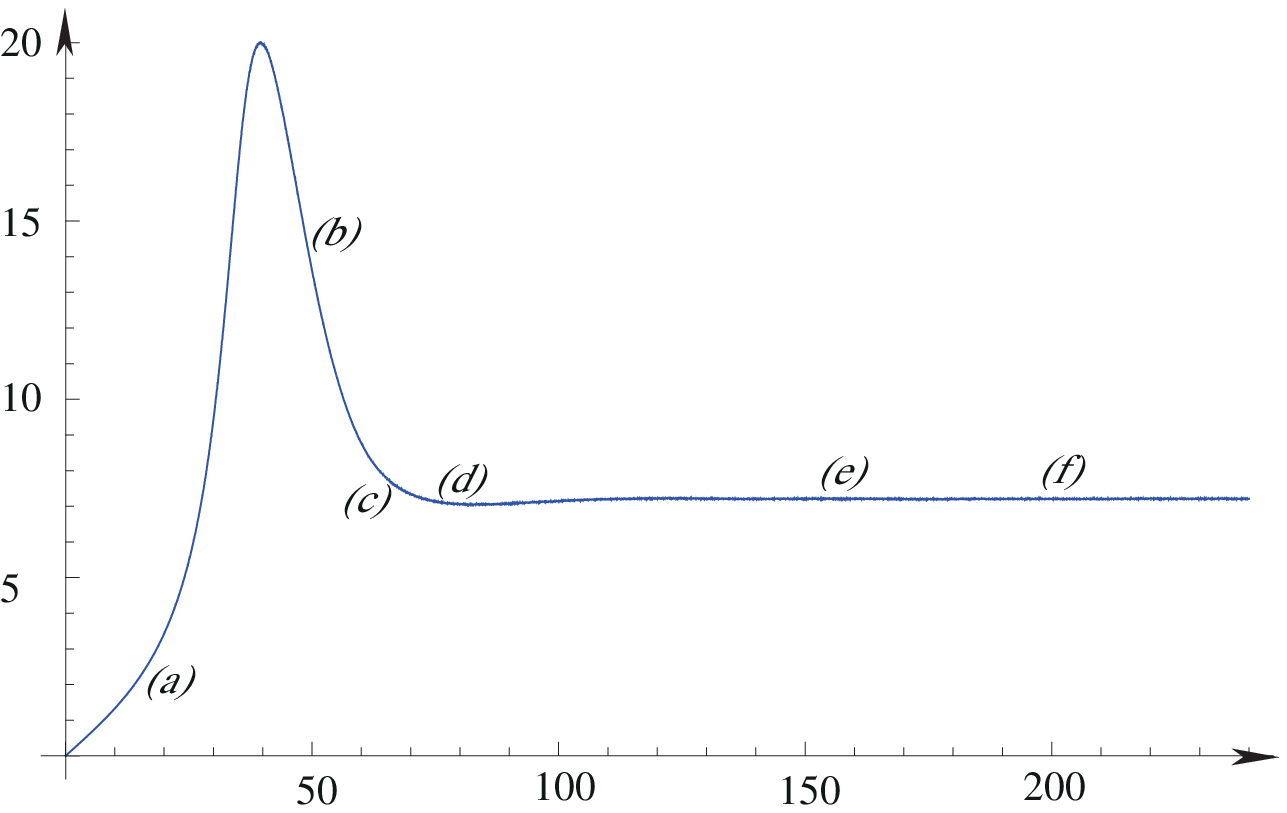} \quad b)\, \includegraphics[width=8cm]{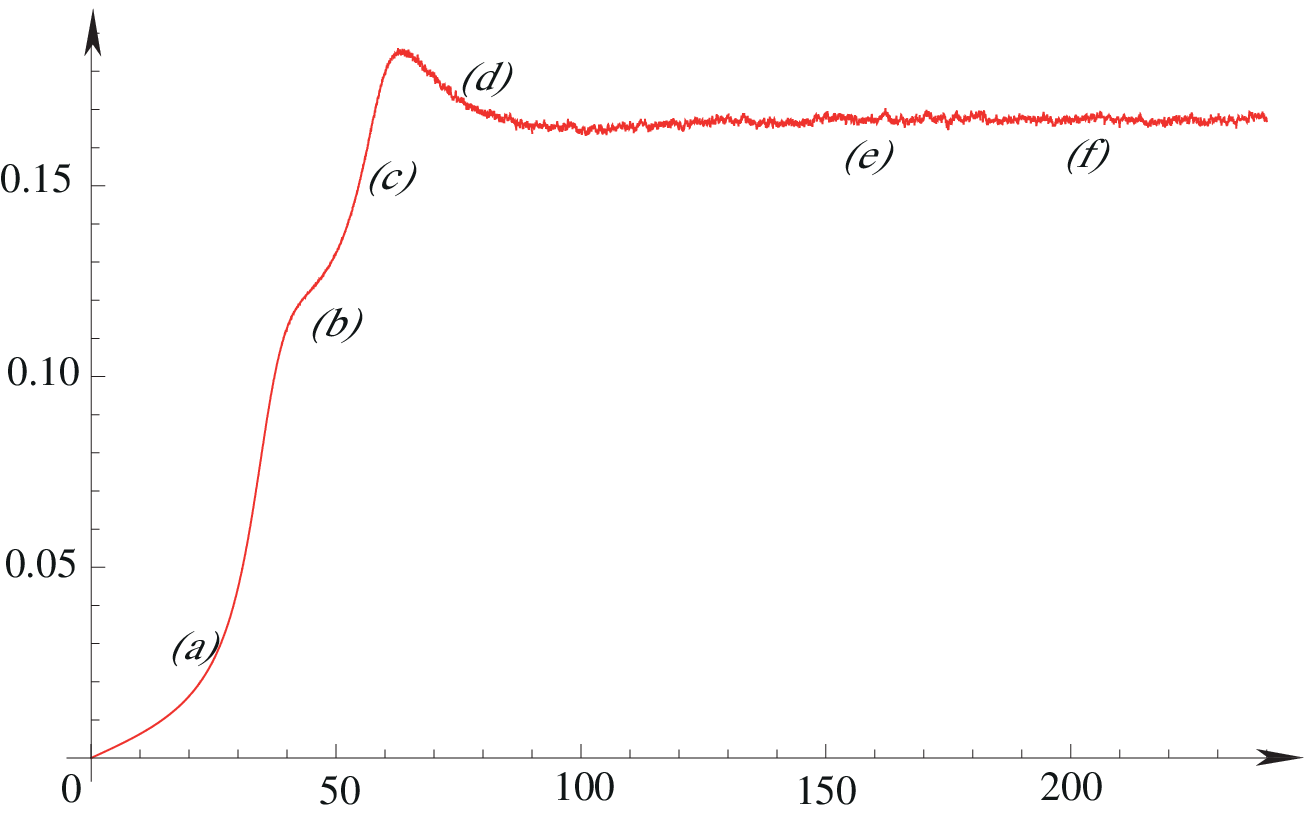} }
\end{center}
\begin{picture}(0,0)(10,10)
\put(45,190) {{${\mathcal N}/L^2[\times 10^{-3}]$}}
\put(-215,190) {{${\mathcal E}/L^2[\times 10^{-3}]$}}
\put(-30,43) {{$t[\times 10^{3}]$}}
\put(225,43) {{$t[\times 10^{3}]$}}
\end{picture}
\caption{a) Evolution of the energy density ${\mathcal E}/L^2$;  and, b) the wave action density ${\mathcal N}/L^2$ with time.
The parameters are the same than those of Fig. \ref{snapshots}. The time of the different snapshots of Fig. \ref{snapshots} are indicated on the curves.}
\label{dynam}
\end{figure}

Fig. \ref{dynam} shows the numerical evolution of the energy and the wave action with time for this numerical simulation. After a transitory regime where both quantities vary, we observe that a quasi-stationary regime is reached above $10^5$ unit time approximately.

\begin{figure}[h!]
\begin{center}
\centerline{a)\,  \includegraphics[width=8cm]{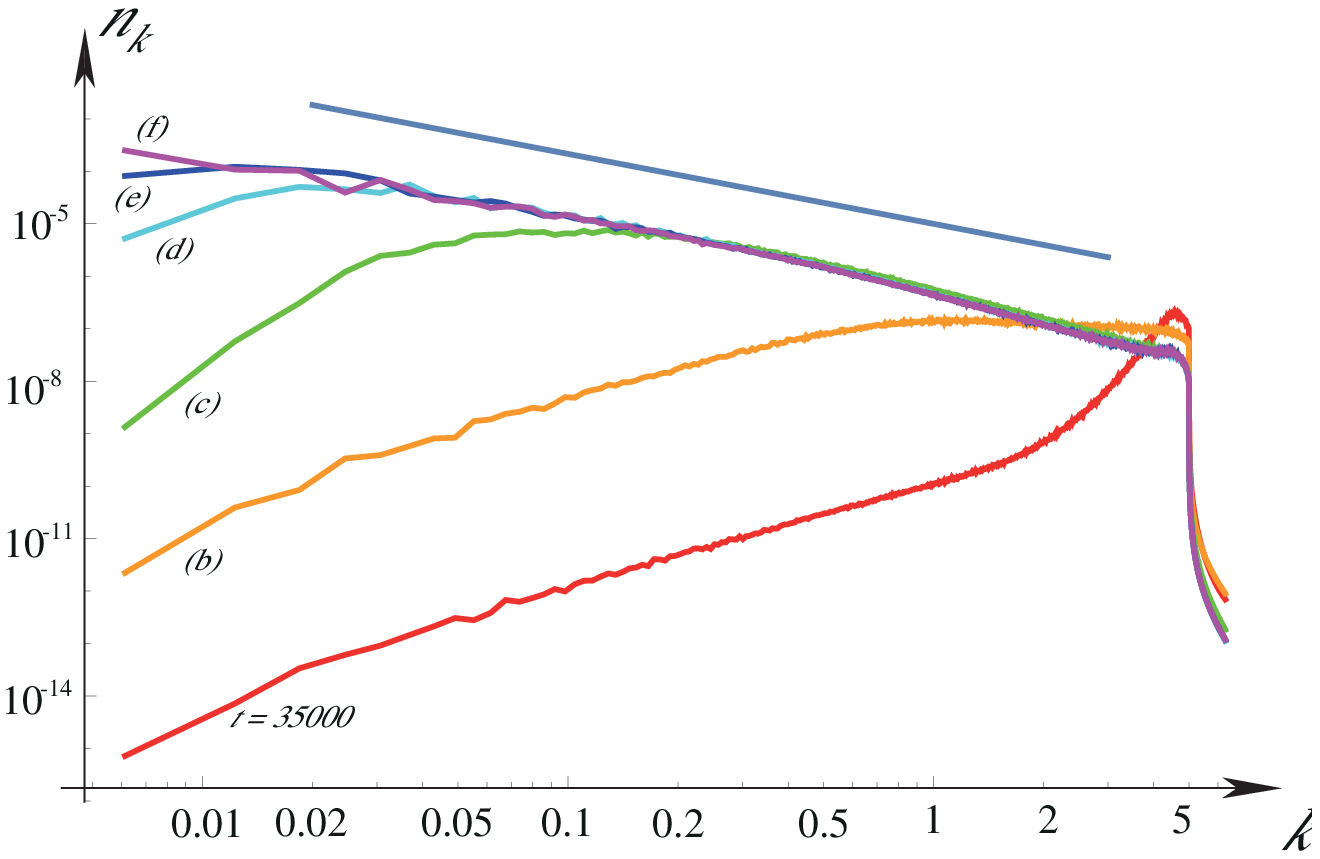}\quad b)\,  \includegraphics[width=8cm]{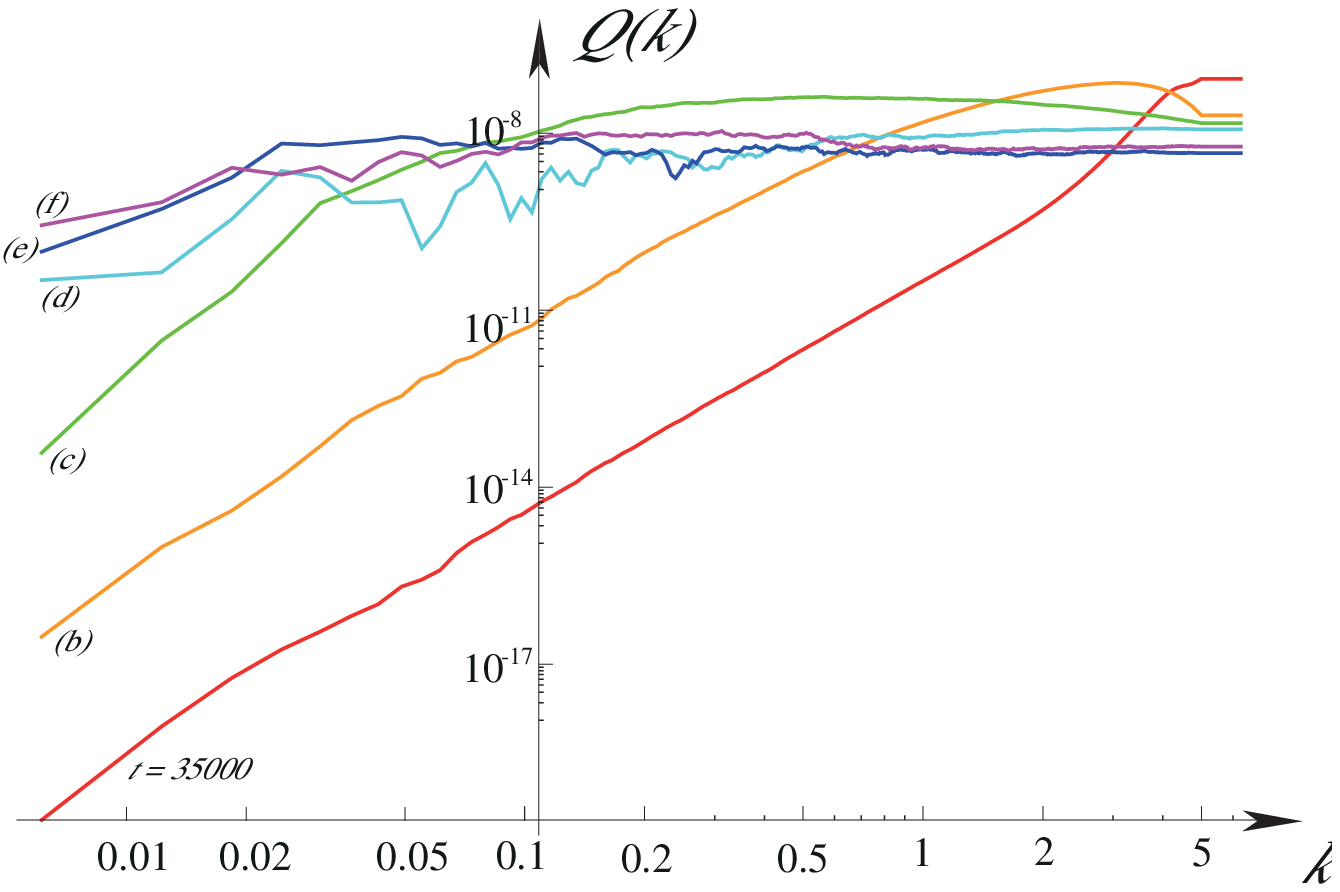} }
\caption{ a) Angular average of the spectra, $n_k$, as a function of wavenumber $k$ in log-log scale, at different time steps starting at  $t=35\times 10^3$. Subsequent spectra are labeled according to the snapshots of Fig. \ref{snapshots}.  The straight line indicates a power law $k^{-4/3}$ as a reference guide.  b) Under  the same conditions log-log plot of the wave action flux $Q(k,t)$ as a function of wavenumber for the same times.}
\label{fig:spec-inv}
\end{center}
\end{figure}

This dynamics can also be inspected within the evolution of the wave spectrum which is defined following eq. (\ref{Canonical}) and (\ref{spectrum}) by:
$$n_{  k}(t) = \omega_k \langle |\zeta_{\bm k}|^2\rangle.$$
 Fig. \ref{fig:spec-inv}-a  shows the averaged value  of the wave spectrum over the angle in the $k$-space.The forcing creates a wave action flux towards $k=0$ that ``fills'' the spectrum at large scale. This can be observed on Fig. \ref{fig:spec-inv}-b which present the wave action flux (\ref{fluxNdef}) at different time, computed explicitly as a sum over discrete modes $Q(k) = 2 \pi \sum_{q=0}^k k\, \partial_t n_k $. (Note that by definition $Q(0)=0$.)
For large time (again, above $10^5$ time units) the spectrum tends asymptotically to a stationary form that exhibits a power law with an exponent surprisingly  close to the hypothetical $4/3$-inverse cascade exponent (\ref{KZinverse}) which is forbidden by the ${3\leftrightarrow1}$ interactions. Similarly, the wave action flux $Q$ converge to an almost constant value. Notice however, that the dynamics is not steady but only in a ``quasi"-stationary regime since the 
large scale modes still exhibit a slow dynamics.

In the following, we will show that the evolution of this amplitude spectra can be decomposed in two distinct  regimes in time, both being dominated by weakly random waves. The first regime displays a self-similar behavior corresponding to a non-constant wave action flux. On the other hand, the later regime displays a quasi-steady behavior consistent with an inverse cascade with a nearly $k^{-4/3}$ spectrum. This regime exhibits an almost constant flux of wave action towards the large scales (see \ref{fig:spec-inv}-b) except precisely near the largest scale of the system ($k\approx 0$).

\section{Signature of a finite-time singularity} \label{sec:blowup} 
The first stage of the evolution appears as the formation in time of a spectrum characterized by a non uniform flux of wave action from the short  to the long scales, 
as shown in figure (\ref{fig:spec-inv}). This flux fills the spectrum from the large $k$ towards small $k$, tending to a 
steady power-law spectrum with an almost constant flux of wave action $Q$ (Fig. \ref{fig:spec-inv}-b). 

It is important to notice that this built in time spectrum is of finite capacity~\cite{newell}, that is $\int_0^k n_k d^2k < \infty$ (taking $n_k \propto k^{-\alpha}$ with $\alpha \sim 4/3$. Therefore one expects,
assuming a constant injection of wave action in the injecting domain around $k_i$, the formation of such spectrum in finite time. This situation is in fact similar to the self-similar formation of a condensate
of weakly classical nonlinear waves~\cite{becnls,becnls2,lacaze,lowtemp} and we shall characterize quantitatively this self-similar dynamics. To do that, we compute the characteristic length scale involved in the self-similar process, via the
negative moments (typically $n \leq -2$ later on) of the spectral distribution~\cite{ZakharovBook}:
\begin{eqnarray} \left< k^n\right> = 
\left(\int k^n n_k (t) \, d^2k \right)/\left(\int  n_k (t) \, d^2k \right). \label{moments}\end{eqnarray}

\begin{figure}[h!]
\begin{center}
\centerline{  \includegraphics[width=8cm]{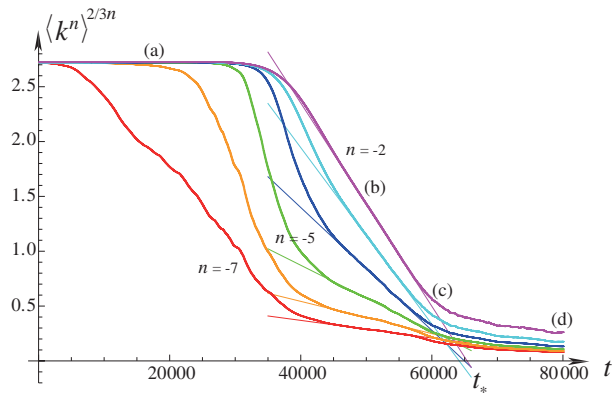} }
\caption{The evolution of the characteristic wave number $[k_0(t)]^{2/3}$ computed through the $n$-th moments  of the distribution (\ref{moments}), for consecutive $n$ ranging from $n=-7 $ up to $n = -2$ (labeled explicitly on the figure). The simulation conditions are as for other figures. The straight lines correspond to a linear fit  $K_n (t_* -t)$. Notice that almost all moments vanish near an unique critical time. The corresponding values for this time are: $t_* = 85971.8$ for $n=-7$, $t_* = 76926.9$ for $n=-6$, $t_* = 69154.8$ for $n=-5$, $t_* = 64971.5$ for $n=-4$, $t_* = 64173.4$ for $n=-3$, $t_* = 65376.9$ for $n=-2$.  Despite  the inaccuracy of the higher order moments (-7,-6) all other critical times  are around $t_*\approx 65000$, indicating the independence of $t_*$ with order $n$ (Notice that the range of the temporal axis is different from the one of Figs. \ref{dynam} and \ref{fig:wNL}).}
\label{fig:moments}
\end{center}
\end{figure}

This allows to define characteristic wave numbers of the spectrum through $ \left< k^n\right>^{1/n}$. Fig. \ref{fig:moments} shows these characteristic wave number 
to the power $2/3$ computed numerically for different moments from $n=-2$ to $n=-7$ at short times ($t<80000$ time units). We observe that the different curves 
exhibit a linear decrease below a critical time $t_*$ suggesting the singular behavior for the critical wavenumber of the spectrum:
\begin{eqnarray}k_0(t) \sim (t_*-t)^{3/2}, \quad {\rm with}\quad t_*\approx 65000.\label{eqn:k0vst}\end{eqnarray}
It is the signature of a finite time singularity that would be present if the asymptotic spectrum would be filled with a constant wave action flux.

This singular behavior suggests a self-similar solution of the form~\cite{lacaze,lowtemp}:
\begin{equation}n_k (t)= \frac{1}{(t_*-t)^{\alpha}}\phi\left( \frac{k}{(t_*-t)^{\beta}} , \log(t_*-t) \right) .\label{selfeq}\end{equation}

From relation (\ref{eqn:k0vst}), we obtain $\beta =3/2$. The parameter $\alpha$ is settled assuming that wave turbulence theory is valid, so that the 
self-similar solution (\ref{selfeq}) should obey the kinetic equation (\ref{KineticEqn}). $\alpha =2$ is then the only possible choice to balance the l.h.s. and the r.h.s terms in the kinetic equation (\ref{KineticEqn}). Finally, the function $\phi$ satisfies an autonomous equation, which reads:
\begin{equation}
\frac{\partial}{\partial\tau}\phi(s,\tau) = \left(2 \phi(s,\tau) +\frac{3}{2} s \frac{\partial}{\partial s}\phi(s,\tau) \right)   -{\mathcal C}[\phi(s,\tau) ]  
\label{SelfKineticEqn}
\end{equation} 
where $s= k (t_*-t)^{-3/2}$ is the self similar variable, $\tau =  \log(t_*-t) $, and $ {\mathcal C}$ is formally the same collisional operator of (\ref{KineticEqn}), but with the scaling variable $s$ instead of $k$. 

\begin{figure}[h!]
\begin{center}
\centerline{  \includegraphics[width=8cm]{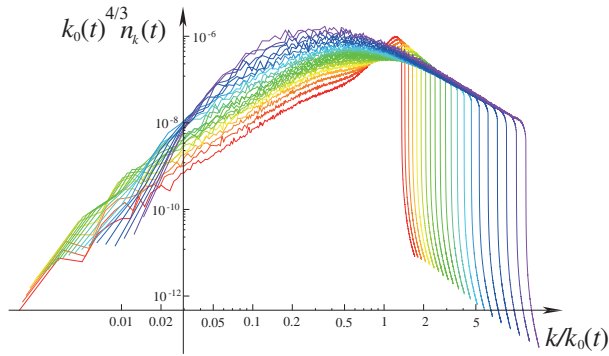} }
\caption{Plot (log-log scale) of the rescaled spectra $k_0(t)^{4/3} n_k(t)$ as function of the rescaled variable $s= k/k_0(t)$, with $k_0(t) = \left< k^{-2}\right>^{-1/2}$. This choice shows a very good collapse of all curves in the large $k$ limit.  The spectra plotted correspond to times ranging from $t=40000$ (colored in red) up to $t=60000$ (colored in violet) units. }
\label{fig:initself-similare}
\end{center}
\end{figure}

Thus the self-similar function $\phi(s,\tau)$ follows an integro-differential equation (\ref{SelfKineticEqn}),  with the boundary condition at the origin, $\phi(0,\tau)  =0$, and with the asymptotic behavior 
\begin{equation}
 \phi(s,\tau) = \frac{1}{s^{2\nu} } e^{\lambda \tau}
\label{SelfKineticEqn2}
\end{equation} 
for  $s\to\infty$ and $\tau \to -\infty$ ($t\to t_*$). The condition $\lambda =2-3\nu$ ensures that, in this limit, the tail of the spectrum does not depend on time, as is observed in Fig. \ref{fig:initself-similare}. Notice that one can always re-scale $\phi$ of the nonlinear equation (\ref{SelfKineticEqn}) to settle the pre factor in (\ref{SelfKineticEqn2}) to unity. 
 
Equation (\ref{SelfKineticEqn}) represents in fact a nonlinear eigenvalue problem for $\nu$, which indicates the power law of the spectrum at large wavenumber. Such problems are difficult 
to solve analytically and even numerically since no systematic approaches exist \cite{lacaze,lowtemp}. Here we will develop an indirect method providing an approximate value only for $\nu$.  

Using the relation between the theoretical values of $\alpha=2$ and $\beta=3/2$, we can rescale
the spectra at different times following the self-similar formula (\ref{selfeq}) by plotting $k_0(t)^{4/3} 
n_k(t)$ as a function of $s= k/k_0(t)$ taking $k_0(t)= \left< k^{-2}\right>^{-1/2}$. A good collapse
of the spectra into a single universal curve is then observed in Fig. \ref{fig:initself-similare} particularly at large wavenumber $s$. 
In fact, the envelope of all the rescaled curves defines the function 
$$\phi_\infty(s)= \lim_{\tau \rightarrow -\infty} \phi(s,\tau).$$

\begin{figure}[h]
\begin{center}
\centerline{a)\, \includegraphics[width=8cm]{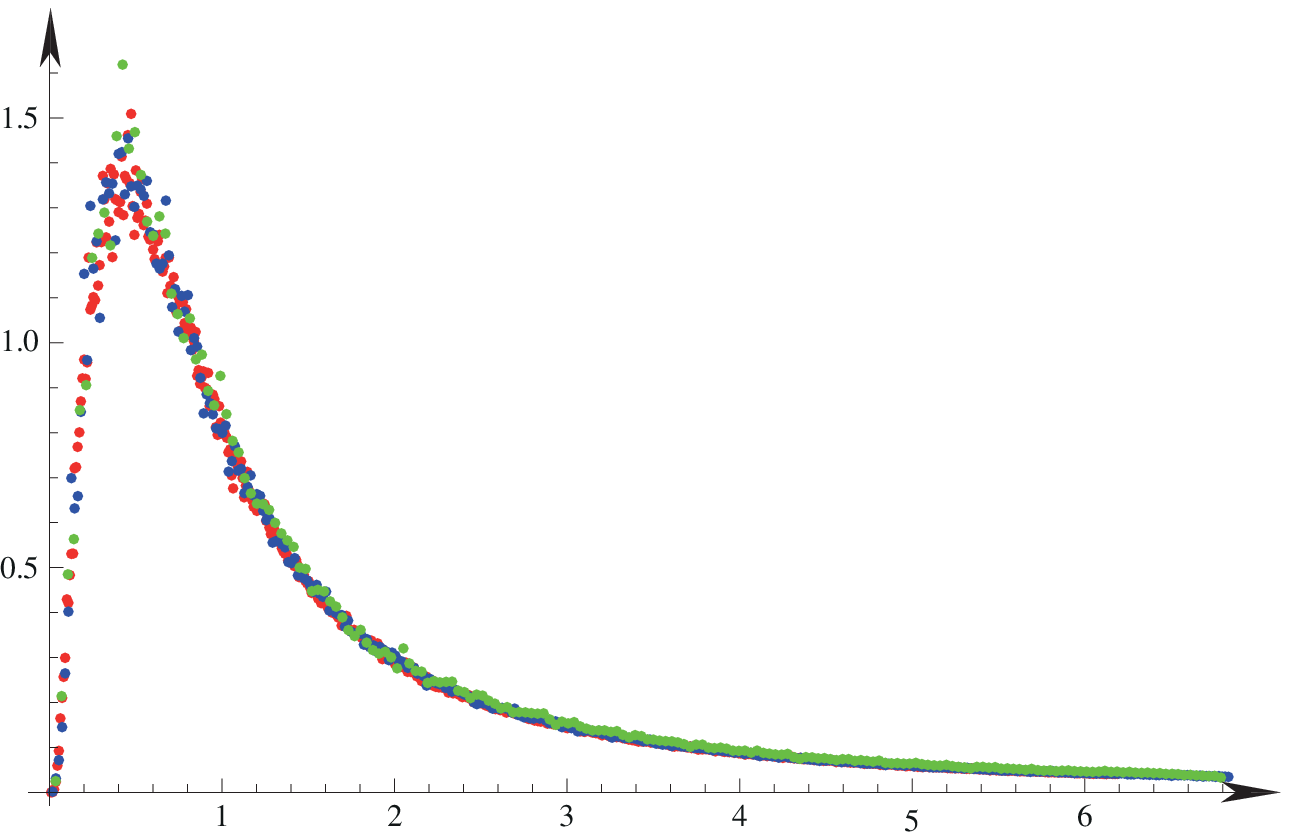} \quad b)\, \includegraphics[width=8cm]{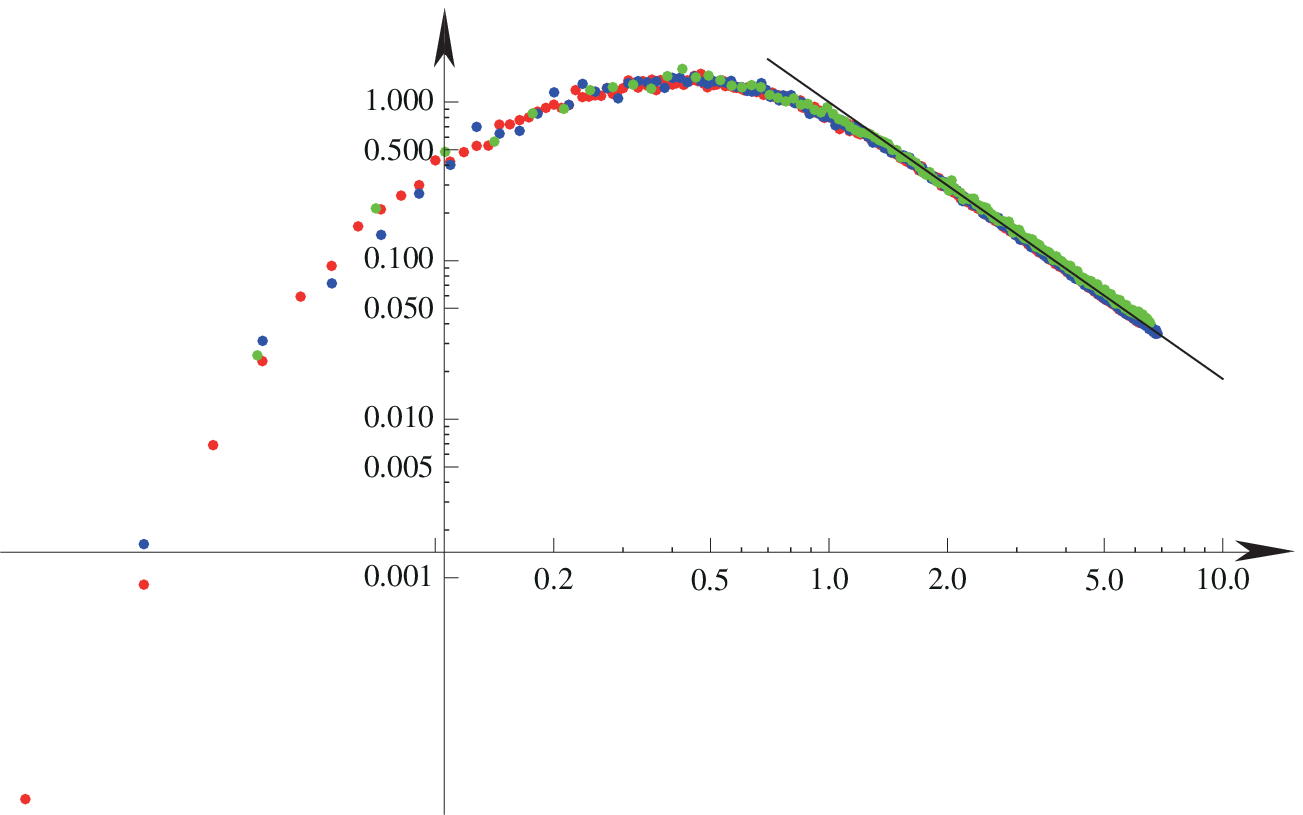} }
\end{center}
\begin{picture}(0,0)(10,10) \put(105,190) {\large{$ \phi_\infty(s)$}} \put(-215,190) {\large{$\phi_\infty(s)$}} \put(-10,40) {\large{$s$}} \put(250,80) {\large{$s$}} \end{picture}
\caption{a) Self-similar universal function $\phi_\infty(s)$ as a function of the self similar variable, $s$, in linear scale. The data comes from the same conditions as Fig. \ref{fig:spec-inv}, but for three distinct system sizes: $L= 256$, green dots; $L= 512$, blue dots; and $L= 1024$, red dots.  b)  The same self-similar universal function, but in log-log scale. Fitting the exponent $\nu$ such that $ s^{2\nu}  \phi_\infty(s)\to 1$ for large $s$ provides $\nu \approx 0.873$ which is slightly larger than $2/3$. }
\label{universalphi}
\end{figure}

Averaging the curves near $t_*$ for the same simulations but with different system sizes $L=256$, $L=512$ and $L=1024$, we obtain a single curve with better resolution for 
$\phi_\infty(s)$, as shown on Fig. \ref{universalphi})-a). Then, seeking the exponent $\nu$ such that $ s^{2\nu}  \phi_\infty(s)\to 1$ for large $s$, the best fit gives
$\nu \approx 0.873$, which is significantly different (higher) than the theoretical value $\nu=2/3$ of the inverse cascade eq. (\ref{KZinverse}). Let us emphasize that it is in fact consistent with such unsteady regime which fills the spectrum from small to large scales.

The particular shape of the universal function $\phi_\infty(s)$ requires a few comments. Firstly, the spectrum decreases near $s=0$, in agreement with  the boundary condition $\phi(0,\tau)  =0$. 
Second, as expected, in the self similar variables the forcing position in the spectrum tends to $s\to\infty$, as one approaches the singularity, therefore the forcing only acts as a boundary condition in the ultraviolet regime. Finally, the matching region between the inner and outer behavior corresponds to the maximum of the function.

\section{Validity of wave turbulence and the late stage regime}

The self-similar behavior (\ref{selfeq}), discussed in the previous section, predicts that wave turbulence assumptions will not be valid near the finite singularity.
In general, wave turbulence theory is no longer valid either because high amplitudes of the spectrum are reached at large scale or because of the discrete dynamics of the modes corresponding to wave lengths close to the size of the computational domain. 

The nonlinear transition due to high amplitude will appear at small $k$ when, for some wave numbers, the nonlinear time scale deduced from 
equation (\ref{KineticEqn}) is of the same order as the period of  the linear wave~\cite{newell}. In the present case of (\ref{selfeq}) one has that the nonlinear and the linear frequencies scale respectively as:
\begin{equation}
\omega_{NL}(k) \sim \frac{1}{n_k} \frac{d n_k}{dt} \sim \frac{1}{t_*-t}\quad {\rm and}\quad \omega_k \sim (t_*-t)^{3},\label{BreakdownCriteria}
\end{equation}
near $t_*$. Therefore one expects the nonlinearities to be large as $t\to t_*$ so that WWT cannot applied anymore.

\begin{figure}[h!]
\begin{center}
\centerline{ a) \includegraphics[width=8cm]{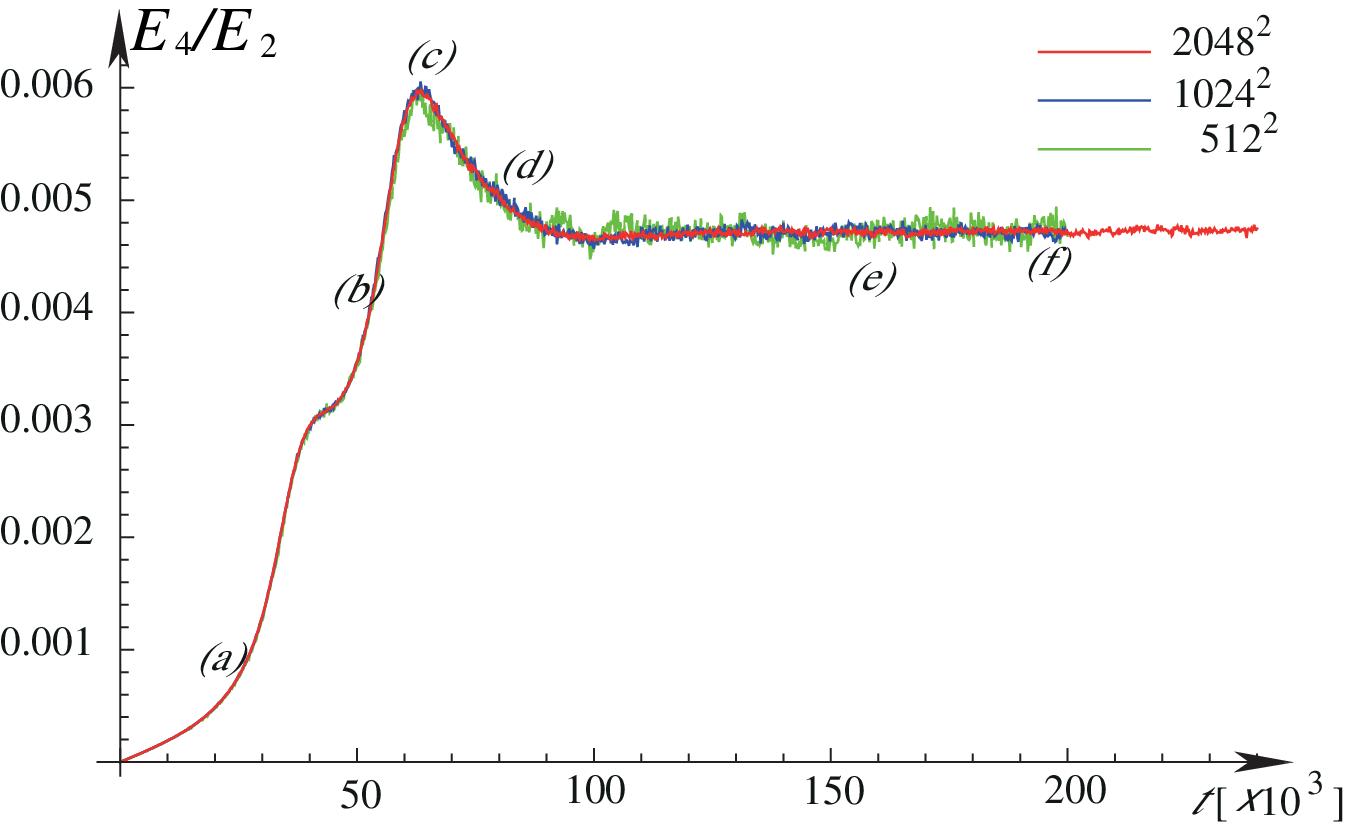}  \quad b) \includegraphics[width=8cm]{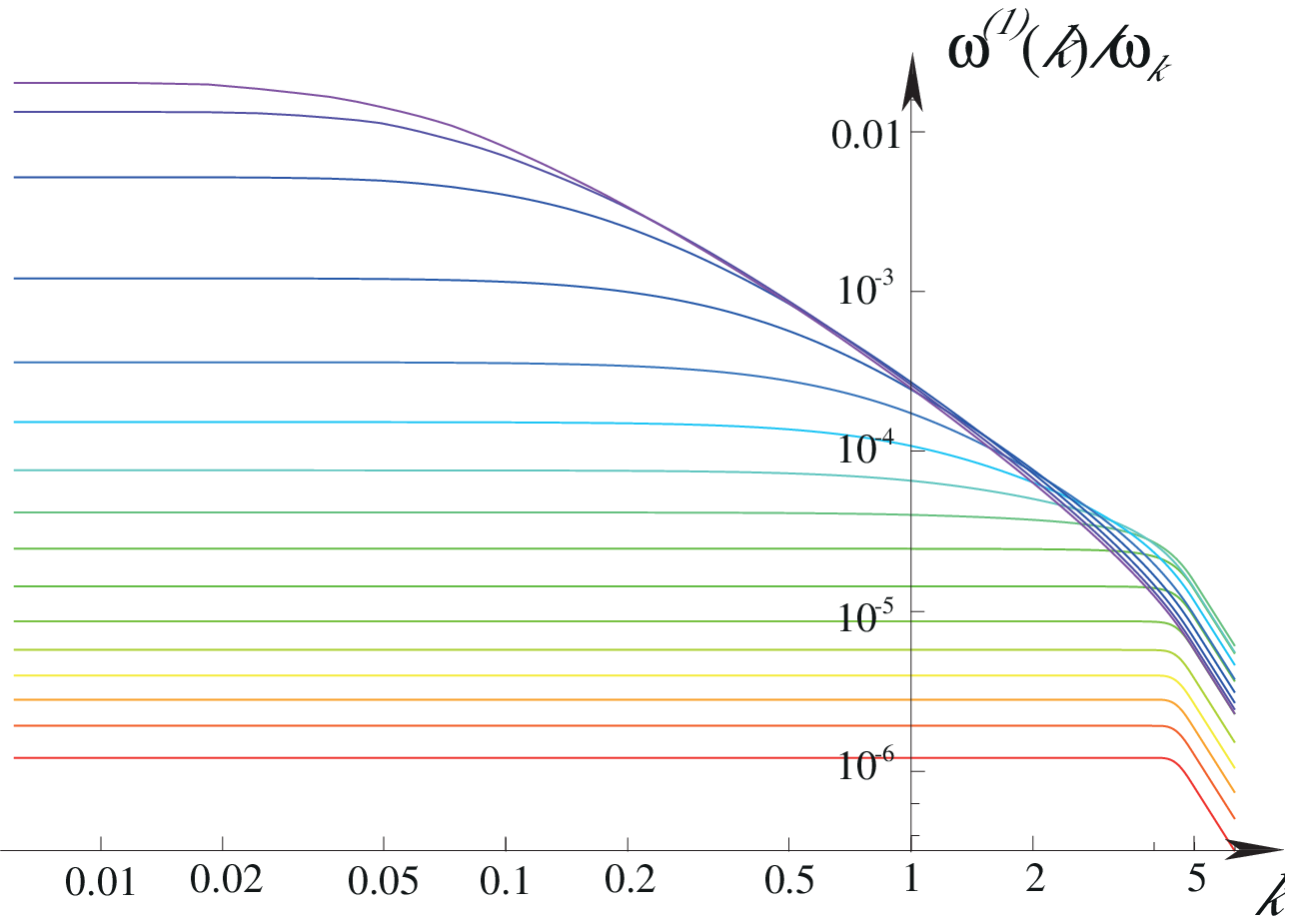}  }
\caption{ a) The ratio $E_4/E_2$ vs. $t$, under the same conditions  as in Fig. \ref{fig:spec-inv} but for three distinct system sizes $L=256,\, 512$ and $1024$ with $512^2$, $ 1024^2$ and $2048^2$ modes respectively.  b) The ratio ${\omega^{(1)}(k)}/{\omega_k}$ as a function of the wave number $k$  for different times steps in log-log scale. The range of time plotted corresponds exactly to the one used in Fig. \ref{fig:initself-similare}, that is from $t = 40000$ time units (colored in red) up to $t = 60000$ (colored in violet). Both quantities show the weakness of the nonlinearities in all wave numbers and at all times, justifying the validity of the weak turbulence theory.  }
\label{fig:wNL}
\end{center}
\end{figure}

In the following we compute numerically two distinct criteria to quantify the ratio between nonlinear and linear contributions.
 First, we computed the ratio between the nonlinear energy, $E_4 = -\int \left(  \frac{1}{2}(\Delta\chi)^2 + \frac{1}{2} \chi \{\zeta,\zeta\} \right)\,dx\,dy\, $, and the linear energy
$E_2 = \int \left(\frac{1}{2}(\partial_t\zeta)^2+ \frac{1}{8} (\Delta \zeta )^2 \right)\,dx\,dy\, $, already discussed in (\ref{totalenergy}). This is a global criteria which 
depends only on time and indicates the relative importance of both energies in the dynamics. Fig. \ref{fig:wNL}a) shows this ratio as a function of time. It is 
observed that $E_4/E_2$ is at most of the order of $6\times 10^{-3}$, indicating that the non-linear contributions (the stretching contributions) to the energy are really small. 
Incidentally, the maximum of  $E_4/E_2$ arises for $t\approx t_*$ confirming the existence of a precursor to a singularity. This can be understood by the following scaling 
argument:  near the 
singularity, the quadratic energy would scale like $E_2 \sim \int \omega_k n_k \,d^2k \sim (t_*-t)^4$, while the fourth order energy (the stretching) would follow
$E_4 \sim \int n_{k} ^2 \, d^2k\sim  1/(t_*-t)$, hence  $E_4/E_2\sim (t_*-t)^{-5}.$

Although this criterium suggests that the nonlinear behavior is globally weak, one cannot ensure 
that the nonlinearities are uniformly weak and in particular that the nonlinearities are 
effectively small for all scales. A local (in $k$) criteria can be used numerically via the ratio between the linear time scale and the first order nonlinear correction to the frequency (\ref{OmegaRenorm}), as stated in eq. (\ref{BreakdownCriteria}).

Fig. \ref{fig:wNL}b) plots the ratio ${\omega^{(1)}(k)}/{\omega_k}$, from (\ref{OmegaRenorm}), as a function of the wave number at different times. One notices that  the infra-red behavior of the quotient ${\omega^{(1)}(k)}/{\omega_k}$ increases significantly when approaching the singularity. Nevertheless, it is always less than $10^{-2}$, which implies that the weak amplitude expansion is presumably uniformly valid, even near the singularity signature.

Though wave turbulence theory predicts breakdown of the theory (\ref{BreakdownCriteria}), 
because $\omega_{NL}(k)/\omega_k\sim (t_*-t)^{-4}$, one observes that direct numerical 
simulations on the F\"oppl--von K\'arm\'an equations (\ref{foppl0},\ref{foppl1}) do not allow the 
nonlinearities to be of the order of unity. Nevertheless, in the case of a large forcing, such effects 
have been seen in our numerics and it could be then a reason for the breakdown of wave turbulence.

In conclusion, in the limit of small forcing investigated here, the system does not create strong nonlinearities although the singular behavior is cured near $t_*$. This effect comes from the discrete properties of the system which becomes relevant at this stage, in particular,  to the discrete dynamics of the first modes of the plate (the lowest in term of frequency) which have to be considered in a modified picture of wave turbulence theory~\cite{karta,shrira}.


\section{Discussion} 

Our numerical study reveals that a wave action inverse cascade is built in time and eventually 
reaches the infrared region in finite time through a clearly identifiable self similar process. In this 
early time regime, the wave system is driven by the WWT kinetic equation (\ref{KineticEqn}), 
and the 
dynamics is characterized by a self similar evolution which should eventually blow-up in finite time.  
However, near the singularity, the dynamics is smoothed and the kinetic equation is no longer 
valid: in the small forcing cases, investigated here, the system is governed by the discrete 
dynamics of the largest modes coupled with the continuous spectrum (a discrete breakdown). On the other hand, for larger forcing (not studied here), a regularization of the dynamics through the nonlinear breakdown of the
 WWT is expected  (nonlinear breakdown).

\begin{figure}[h!]
\begin{center}
\centerline{ a) \,  \includegraphics[width=8cm]{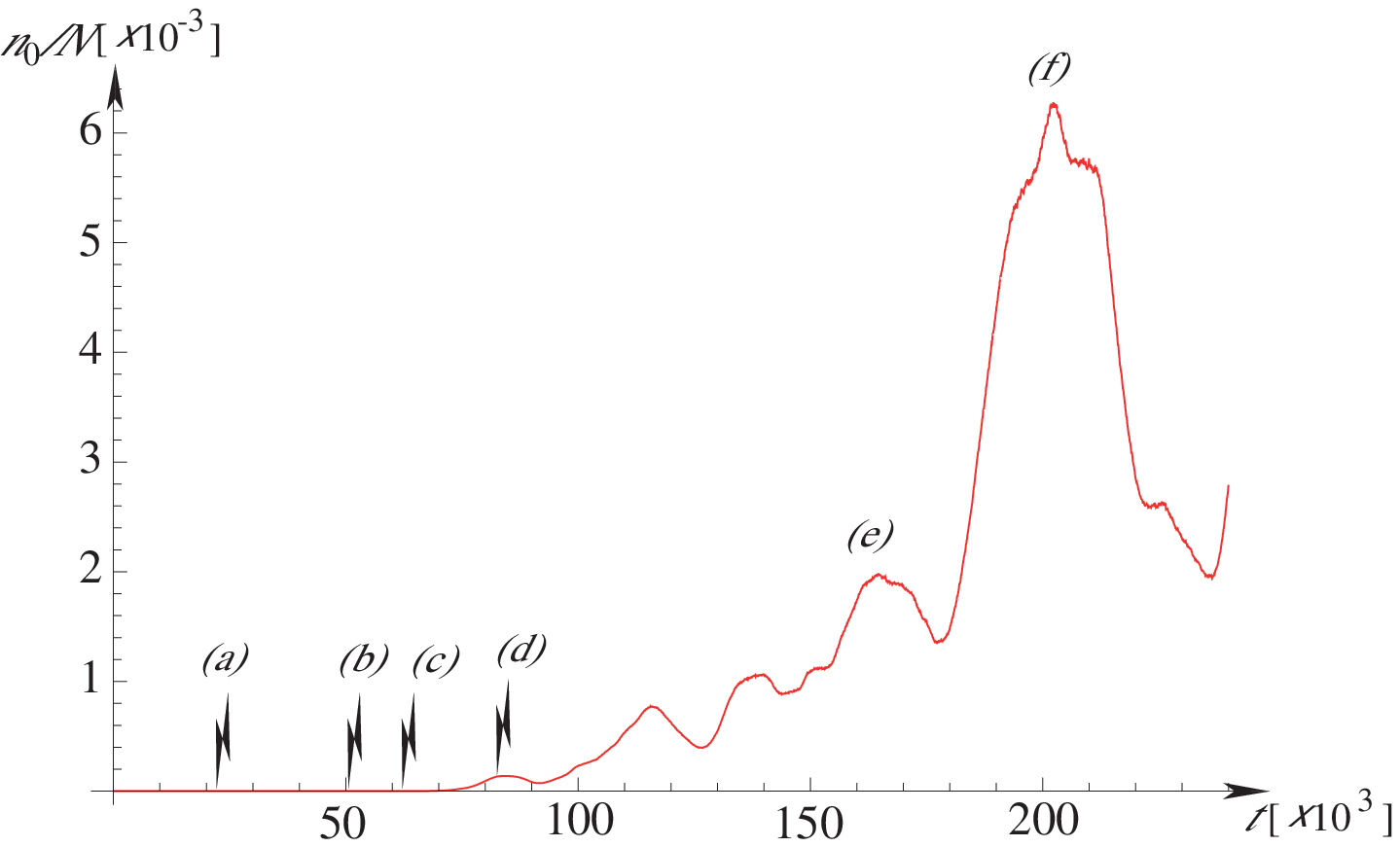}\quad b)\, \includegraphics[width=8cm]{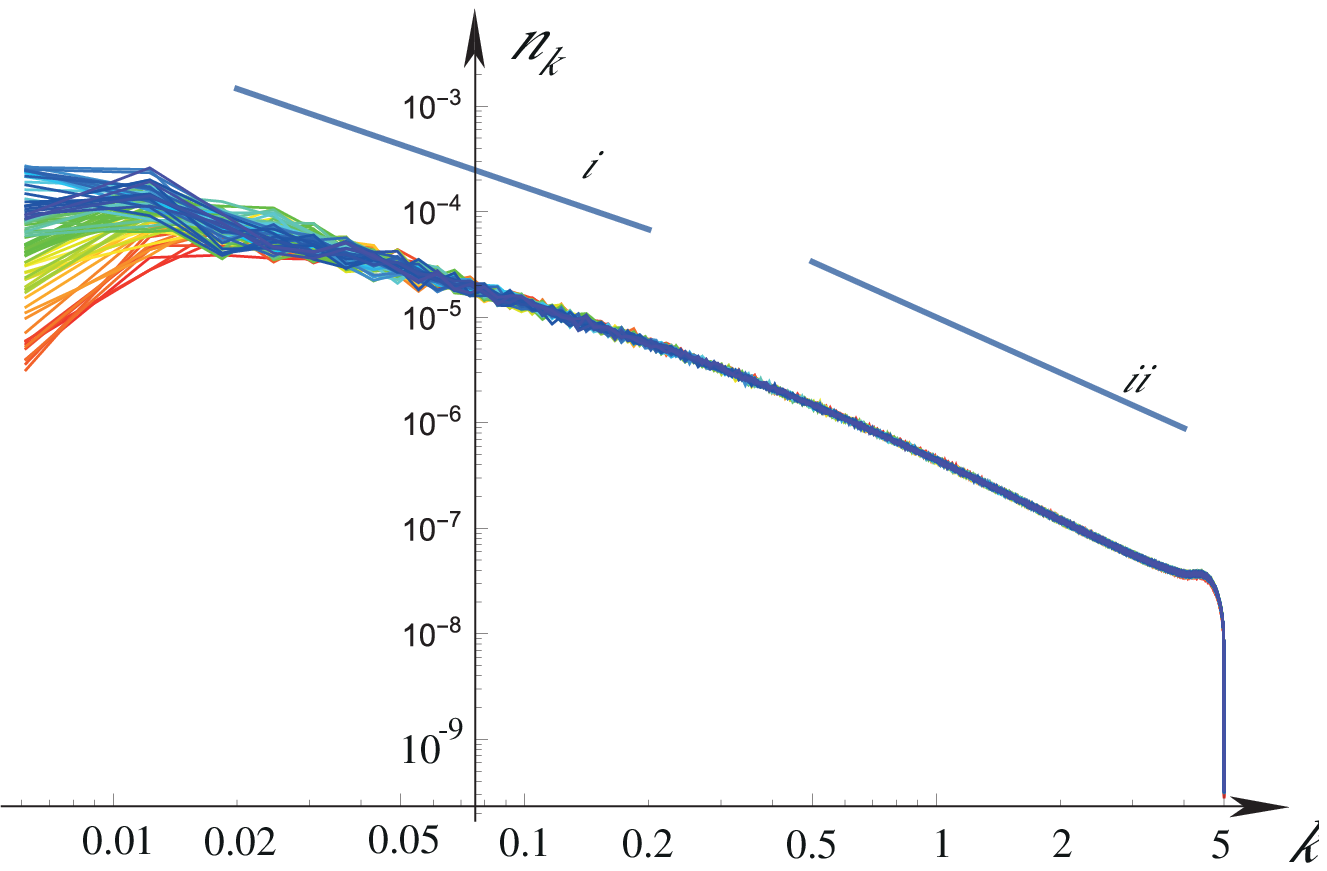}  }
\caption{ a)  Evolution of the fraction of the first mode. This plot corresponds exactly to the numerical simulation of Fig. \ref{snapshots} and the times of the different snapshots are indicated. b) The long-time evolution of the spectra. The time considered are from $t=80000$  units (colored by red) up to $t=240000$ (colored by violet). The spectra only vary for the small wavenumbers where it relaxes towards the stationary power law spectrum $n_k \propto k^{-2 \nu}$ with $\nu \sim 2/3$. The line ({\it i}) correspond to a 4/3-power law, while the line ({\it ii}) corresponds to a $2\times 0.873$ power law which represents better the behavior in the large $k$ limit.}
\label{fig:condensate}
\end{center}
\end{figure}

For other systems where the wave action is a conserved quantity (for instance for the nonlinear Schr\"odingier equation),  a condensate at (or around) $k=0$ forms, changing the post blow-up dynamics~\cite{becnls,becnls2}. Such effects are not possible here since the wave action is not conserved by the dynamics. The mode $k=0$ is neutral and remains null with time. Nevertheless, as it has been shown for the non-linear Schr\"{o}dinger equation~\cite{falko13}, the first modes of the systems can exhibit an autonomous dynamics. Figure \ref{fig:condensate}-a) shows precisely the evolution with time of the amplitude for the lowest mode of the plates: after the blow-up time $t^*$, the amplitude of this mode grows more or less linearly in time, but exhibiting also important oscillations. Finally, for these large times, a stationary regime is eventually reached where the spectrum behaves approximatively like $n_k \sim 1/k^{4/3}$ for the low wave numbers ($k<0.2$ in Fig. \ref{fig:condensate}-b) and  $n_k \sim 1/k^{2\times 0.873 }$ for the smaller wavelength ($0.5< k< k_d$ in Fig. \ref{fig:condensate}-b). This stationary regime is not surprising, since we expect that the long time behavior does not change the power law built by the self similar evolution (Section \ref{sec:inverse}), and long wave modulations transfer wave action toward  $k=0$ with precisely the Kolmogorov-Zakharov spectrum, $n_k \sim 1/k^{4/3}$, which corresponds to the constant 
wave action flux solution found for ${\mathcal C}_{22}[n_k]$. However, notice that this regime is not formally steady for the smallest wave numbers, where the
dynamics appears rather more like a relaxation towards a full stationary dynamics.

In conclusion, although an elastic vibrating plate does not formally posses a wave action conservation law, an undoubtedly inverse cascade of wave action is observed, exhibiting a complex time dependent dynamics. The process of formation of such an inverse cascade is ruled by a self similar evolution of the spectra which transfers wave action from short wavelength scales up to long wave length scales. Formally the observed self similar solution blows-up leading to a singular behavior which is cured in the original system, probably, because of the finite size of the system and the role played by the discreteness of the modes. The late evolution of the system is also governed by wave turbulence theory, although the discrete nature of the lowest modes modifies the overall picture. This scenario is consistent with the formation of a coherent structure which is characterized by the largest modes of oscillation (see Fig. \ref{fig:wNL}a) plus small fluctuations. Remarkably, the nonlinear fraction of the energy indicates that this coherent structure makes the stretching very small. 

The authors would like to thank Peter Mason for numerous comments and fruitful discussions. G.D. acknowledges  support from  CONICYT  PAI/Apoyo al Retorno 82130057, C.J. and S.R. acknowledge the FONDECYT grant N 1130709 (Chile). SR. is on leave from Institut Non Lin\'aire de Nice, UMR 6618 CNRS-UNSA, 1361 Route des Lucioles, 06560 Valbonne, France, EU.

\end{document}